\documentclass[
10pt, 
a4paper, 
oneside, 
headinclude,footinclude, 
BCOR5mm, 
]{scrartcl}

\usepackage[
nochapters, 
beramono, 
eulermath,
pdfspacing, 
dottedtoc 
]{classicthesis} 

\usepackage{arsclassica} 

\usepackage[T1]{fontenc} 

\usepackage[utf8]{inputenc} 

\usepackage{graphicx} 
\graphicspath{{Figures/}} 

\usepackage{enumitem} 

\usepackage{lipsum} 

\usepackage{subfig} 

\usepackage{amsmath,amssymb,amsthm} 

\usepackage{varioref} 

\theoremstyle{definition} 

\theoremstyle{plain} 

\theoremstyle{remark} 

\hypersetup{
colorlinks=true, breaklinks=true, bookmarks=true,bookmarksnumbered,
urlcolor=webbrown, linkcolor=RoyalBlue, citecolor=webgreen, 
pdftitle={}, 
pdfauthor={\textcopyright}, 
pdfsubject={}, 
pdfkeywords={}, 
pdfcreator={pdfLaTeX}, 
pdfproducer={LaTeX with hyperref and ClassicThesis} 
} 

\usepackage{enumitem}

\usepackage{graphicx}
\usepackage{amsmath,amsfonts}
\usepackage{algorithmic}
\usepackage{caption}
\usepackage{comment}
\usepackage{xspace}
\usepackage{multirow}
\usepackage{hhline}
\usepackage{oplotsymbl}
\usepackage{tabularx}
\usepackage{booktabs}
\usepackage{colortbl}

\newcommand{\ptfa}{2D-2FA\xspace}
\newcommand{\stfa}{PIN-2FA\xspace}
\newcommand{\ntfa}{Push-2FA\xspace}
\newcommand{\Us}{\mathcal{U}}
\newcommand{\Sr}{\mathcal{S}}
\newcommand{\Dv}{\mathcal{D}}
\newcommand{\Cl}{\mathcal{C}}
\newcommand{\pw}{\mathsf{pw}}
\newcommand{\pn}{\text{PIN}}
\newcommand{\pt}{\mathsf{id}}

\newif\ifshowcomments
\showcommentsfalse

\ifshowcomments
\newcommand{\maliheh}[1]{{\color{cyan}[Maliheh: #1]}}
\newcommand{\shashank}[1]{{\color{olive}[Shashank: #1]}}
\newcommand{\sha}[1]{\shashank{#1}}
\else
\newcommand{\maliheh}[1]{}
\newcommand{\shashank}[1]{}
\newcommand{\sha}[1]{}
\fi

\newif\ifshownotes
\shownotesfalse

\ifshownotes
\newcommand{\mynote}[1]{{\color{blue}[Note: #1]}}
\else
\newcommand{\mynote}[1]{}
\fi

\newcommand{\adv}{\mathcal{A}}

\newcommand{\user}{\mathcal{U}}

\newcommand{\mypara}[1]{\paragraph{#1}}

\newif\ifnewcomments
\newcommentsfalse

\ifnewcomments
\newcommand{\nmaliheh}[1]{{\color{cyan}[Maliheh: #1]}}
\newcommand{\nsha}[1]{{\color{olive}[Shashank: #1]}}
\else
\newcommand{\nmaliheh}[1]{}
\newcommand{\nsha}[1]{}
\fi

\newif\ifshowhighlights
\showhighlightsfalse

\ifshowhighlights
\newcommand{\hit}[1]{{\color{blue}#1}} 
\else
\newcommand{\hit}[1]{#1}
\fi

\newcommand{\id}{identifier\xspace}
\newcommand{\ids}{identifiers\xspace}

\newcommand{\Id}{Identifier\xspace}
\newcommand{\Ids}{Identifiers\xspace}

\newcommand{\uname}{\mathsf{un}}
\newcommand{\sname}{\mathsf{sn}}

\title{\normalfont\spacedallcaps{\ptfa: A New Dimension in Two-Factor Authentication}}

\author{\spacedlowsmallcaps{Maliheh Shirvanian}\thanks{\texttt{mshirvan@visa.com}. Visa Research, Palo Alto, CA} \spacedlowsmallcaps{ \& Shashank Agrawal}\thanks{\texttt{shashank.agrawal@wdc.com}. Western Digital, Milpitas, CA}} 

\date{}

\begin{document}

\renewcommand{\sectionmark}[1]{\markright{\spacedlowsmallcaps{#1}}} 
\lehead{\mbox{\llap{\small\thepage\kern1em\color{halfgray} \vline}\color{halfgray}\hspace{0.5em}\rightmark\hfil}} 

\pagestyle{scrheadings} 

\maketitle



\section*{Abstract}
We propose a two-factor authentication (2FA) mechanism called \ptfa to address security and usability issues in existing methods. \ptfa has three distinguishing features: First, after a user enters a username and password on a login terminal, a unique \emph{\id} is displayed to her. She \emph{inputs} the same \id on her registered 2FA device, which ensures appropriate engagement in the authentication process. Second, a one-time PIN is computed on the device and \textit{automatically} transferred to the server. Thus, the PIN can have very high entropy, making guessing attacks infeasible. Third, the \id is also incorporated into the PIN computation, which renders \textit{concurrent attacks} ineffective. Third-party services such as push-notification providers and 2FA service providers, do not need to be trusted for the security of the system. \hit{The choice of identifiers depends on the device form factor and the context. Users could choose to draw patterns, capture QR codes, etc.}





\hit{We provide a proof of concept implementation, and evaluate  
performance, accuracy, and usability of the system. 
We show that the system offers a lower error rate (about half) and better efficiency (2-3 times faster) compared to the commonly used PIN-2FA.
Our study indicates a high level of usability with a SUS of 75, and a high perception of efficiency, security, accuracy, and adoptability. }







\newpage


\section{Introduction}
\label{sec:intro}



Two-factor authentication (2FA) is widely used to add an extra layer of security to  services that rely on passwords \cite{two-factor-auth,tnw-auth}. We can broadly classify adopted solutions into two categories, PIN-based and push-based. In PIN-based model, one-time PINs can be delivered to users via SMS or voice, or generated on personal devices \cite{google-authenticator,safenet-mobilepass}. Users are then supposed to copy the PIN to the client terminal (e.g., web browser). Push-based 2FA operates by pushing notifications to users' devices whenever login attempts are made \cite{twilio-push,duo-push} and users can choose to accept or decline the attempt. 

Despite known benefits, PIN- and push-based 2FA have several security and usability issues. \stfa suffers from problems like SIM swap attacks \cite{nist-sms-2fa,duo-2fa-survey}, shoulder surfing \cite{Eiband:2017:USS:3025453.3025636}, short PINs \cite{ndss-2014}, etc. 
\nsha{Replace low user preference with long input time and provide a relevant citation. Couldn't find a reasonable citation}
On the other hand, \ntfa suffers from neglectful user approvals and reliance on push-notification services, which can be targeted \cite{li2014mayhem, loreti2018push,xu2012abusing,pushattack1,pushattack2,abbott2020mandatory}. See Section \ref{sec:related} for a detailed discussion.






In this work, we introduce a new 2FA approach called \textbf{\ptfa}. In \ptfa, when a user enters a username and password on a login terminal, a \emph{unique \id} is displayed to her. The user is supposed to \emph{input} the same \id on her personal device, say $\Dv$, to ensure proper engagement. Then, a one-time PIN is generated \emph{on the device} and transferred \emph{automatically} to the server, say $\Sr$, along with the \id. The \id is also incorporated into the PIN computation, binding the PIN to a specific session. \hit{The choice of identifiers depends on the device's form factor and the context. Inputting identifiers could take the form of drawing patterns, capturing QR codes, copying short strings, etc.}

$\Dv$ and $\Sr$ agree on a secret key during a one-time registration process, and compute the PIN using this key. The PIN computation function is defined as $F_k(x)$, where $F$ is a message authentication code \cite{hmac-rfc}, $k$ is the shared secret key, and $x$ is a value known to both $\Dv$ and $\Sr$ (the \id is a part of $x$). Both the PIN and \id are transferred automatically to $\Sr$ for verification. On receiving them, $\Sr$ authenticates the session associated with the \id by verifying the corresponding PIN. Note that the server takes two \underline{D}imensions (\underline{2D}-2FA), PIN and \id, into account.




\mypara{One-time PIN} Observe that the way one-time PINs are generated and circulated in \ptfa is different from \stfa. In the latter, PINs are generated on the server side and sent to the user device \cite{sms2fa} (or they may be generated on the user device itself \cite{google-authenticator, authy}). Then, the user \emph{manually} copies the PIN from the device to the login terminal. After that, the terminal transfers the PIN to the server. In contrast, \ptfa always generates PINs on the device and \emph{directly} transfers them to the server\textemdash without user involvement. One important benefit of this is that PINs can have high entropy, making guessing attacks infeasible.

\mypara{\Id} \ptfa engages users in the authentication process by asking them to copy an \id from the login terminal to their personal device. After the \id has been entered, a PIN is computed and sent (directly) to the server. This stands in contrast to \ntfa where users are just supposed to choose between approve or reject, and may accidentally approve a concurrent session launched by an attacker. Note that \id memorability and secrecy are not relevant to us because neither do users need to remember \ids nor do they need to keep them secret (\ids just need to be unique for each concurrent session). 

\begin{figure}[t]
\centering
\includegraphics[width=0.95\textwidth]{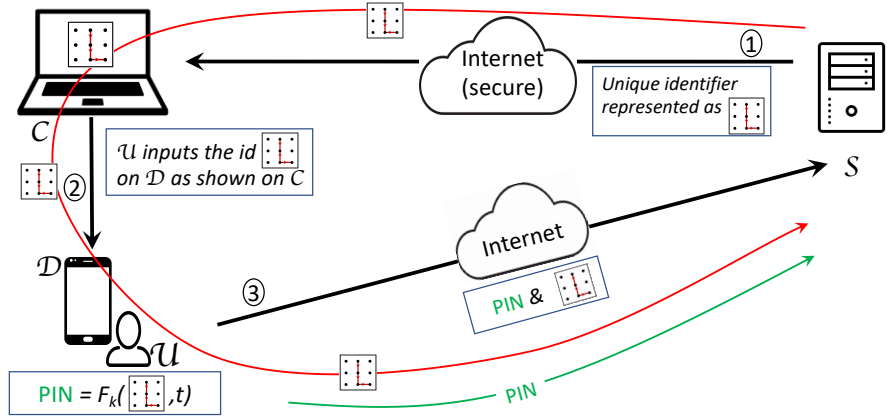}
\vspace{-30mm}
\caption{\ptfa\ system components and interaction model. \hit{\Ids ensure proper user engagement. In this figure, they are instantiated with patterns. User is shown a pattern on the client terminal. She draws the same pattern on her device.} Identifiers are incorporated into PIN computation to bind PINs to sessions. PINs are used to verify availability of the second factor.}
\label{fig:system}
\end{figure}

\mypara{Flow} Notice how the flow of second factor in \ptfa is different from \stfa. In \stfa, a user copies a PIN from the device to the terminal whereas in \ptfa, she copies the \id (not any PIN) in the reverse direction (from terminal to device). Also note that while \ntfa pushes notifications to the device, \ptfa sends PINs to the server. See Figure \ref{fig:system} for an overview.


\mypara{Security} \ptfa\ offers the following security properties: 

\begin{itemize}

\item \textit{Security against client compromise:} An attacker with knowledge of the password cannot authenticate without getting access to the user's  device. Since PIN is transferred automatically, it can have high entropy, making PIN guessing infeasible. 

\item \textit{Security against channel compromise:} Stealing PINs does not help an attacker authenticate to a different session since PIN is a function of both secret key and \id. Thus, PINs can be communicated to the server over an attacker-controlled channel. Shoulder-surfing attacks are not a problem either (as opposed to PIN-based 2FA \cite{Eiband:2017:USS:3025453.3025636}) because \ids are not confidential values.


\item \textit{No reliance on third parties:}
In \ptfa, users do not have to trust third-party services 
such as push-notification providers \cite{google-cloud-messaging,apple-notifications}, 2FA service providers \cite{authy,duo}, etc. 

\item \textit{Resistance to user negligence:} \ptfa\ requires the user to approve the ongoing session by copying a simple \id displayed to her. This level of user engagement is more resistant to user errors as compared to Push-based 2FA, which only requires approval of a notification message \cite{push-accept-dark-side}. Our user study (Section \ref{sec:usability}) shows a 0\% false accept rate indicating that with a more involved user engagement process, erroneous acceptance of an attacker session could be eliminated.

\end{itemize}

\mypara{More features} \ptfa\  offers the following features to enhance  user experience. 

\begin{itemize}

\item \textit{Easy PIN transfer:} Transfer of the PIN is automated and does not require manual copying.




 
\item \textit{Client compatibility:} \ptfa\ does not require any changes on the client, nor any additional software or plugins unlike other approaches such as FIDO2 that requires significant browser support \cite{fidou2f,u2f-not-prime-time,duo-2fa-survey}. 


 
\item \textit{Use of different devices:} \hit{Different user devices come with different input options. While some devices like smartphones have multiple input options (touchscreen, camera, etc.), there are many devices that have only one option. For instance, many smart-home devices, smartwatches, etc.\ just have a touchscreen but no camera to capture QR codes; some devices may only have a small keypad with some buttons but no touchscreen. Fortunately, \ptfa supports multiple identifier types as the \textit{second} dimension like drawing patterns on a touchscreen, capturing QR codes with a camera, typing short strings on a keypad, etc.}


\end{itemize}

\subsubsection*{Contributions} Our contributions can be summarized as follows: 


\mypara{System Design (Section \ref{sec:system}).}
We design a new two-factor authentication mechanism, \ptfa. This design allows automatic transfer of PINs from a device to server with no reliance on third-party services. The PIN is unique for each authentication session 
and corresponds to an \id displayed on the client and copied  by the user to her personal device. 




\mypara{Security Analysis (Section \ref{sec:security-analysis}).} We present a comprehensive security analysis of the system. We consider various types of threats in our analysis. We argue that \ptfa is secure against client compromise because it is infeasible to guess the PIN, even when the attacker tries to log in at the same time as the right user (they receive different \ids). We also show that eavesdropping \ids or PINs on the communication channels is of no use because PINs are a function of both time/counter and \id. We consider several other attack venues too; see section \ref{sec:security-analysis} for more details.

We also compare \ptfa with other well-known 2FA schemes in the framework of Bonneau et al.\ \cite{bonneau2012quest}. Unlike most other schemes, \ptfa is resistant to physical and internal observation.




\mypara{System Implementation (Section \ref{sec:deploy}).} 
We discuss the choice of \ids that can improve the security,  accuracy, and usability of \ptfa. 
We suggest two forms of \ids: 1) PT-2D-2FA, in which \id is represented as a pattern on a $3 \times 3$ grid of dots, similar to Android pattern lock; and 
2) QR-2D-2FA, in which \id is encoded into a QR Code and can be scanned through any device with a camera. 
We develop a full proof of concept implementation of \ptfa, including server-side scripts and an Android app.

\mypara{Evaluation and User Study (Section \ref{sec:usability}).} 
We ran a user study with 30 participants and  evaluated  performance, accuracy, and usability of the system,  and 
compared our methods with the commonly used \stfa.
The study shows that the false reject rate of the system is low (about 2\%) and the false accept rate is 0\%.
The system also seems to be efficient with an end-to-end delay of less than 5s. 
Answers to the usability questions indicate that \ptfa has ``Good'' usability with a system usability scale (SUS) of 75.
The participants were in agreement that the system is efficient, secure, and has low error rate. 
In comparison to \stfa, the system provides similar or better usability rating, which considering the long history of \stfa adoption and familiarity of the users with this technique, is a promising outcome.
2D-2FA error rate was about half that of \stfa and the system end-to-end delay was 2-3 times less than \stfa.
Besides, 75\% of the participants preferred at least one of the 2D-2FA variants (pattern or QR Code) over \stfa.

\phantom{} 
 

\section{Background and Related Work}
\label{sec:background}

\subsection{Notation and Primitives}

We use calligraphic math fonts to denote various entities involved in authentication like $\Us$ for user, $\Cl$ for client, $\Dv$ for device, $\Sr$ for server, etc. 

\mypara{Channels} We use $\mathcal{A} \rightarrow \mathcal{B}$ to denote a channel from $\mathcal{A}$ to $\mathcal{B}$. $\mathcal{A}$ can use this channel to send messages to $\mathcal{B}$ in a reliable and ordered manner (e.g.,\ TCP).
We use $\mathcal{A} \Rightarrow \mathcal{B}$ to denote a \emph{secure} channel from $\mathcal{A}$ to $\mathcal{B}$. A secure channel provides confidentiality and strong integrity protection (e.g.,\ TLS). Messages sent on the channel are hidden from other parties and cannot be modified. (Some metadata like message length, delivery time, etc.\ could be leaked.)

A secure channel from $\mathcal{A}$ to $\mathcal{B}$ does \emph{not} imply a secure channel (or even a regular channel) from $\mathcal{B}$ to $\mathcal{A}$. When there \emph{is} a secure channel available in both directions, we denote it by ``$\Leftrightarrow$''. On the internet, TLS is widely used to establish bi-directional secure channels between clients and servers. Moreover, servers often provide certificates to identify themselves to the clients.


\mypara{Message Authentication} Hash-based message authentication code (HMAC) is a mechanism for message authentication using cryptographic hash functions like SHA-256 or SHA-3 \cite{hmac-rfc}. It takes a key $k$ and a message $m$ as inputs, and outputs a tag. HMAC provides both integrity and authenticity. Without knowledge of $k$, it is infeasible for an attacker to produce a valid (message, tag) pair.  

\mypara{Digital signature} Digital signature is the public-key analogue of MAC. A secret key $sk$ is used to sign messages, which can be verified by anyone with the corresponding public key $pk$. Without knowledge of $sk$, it is infeasible to forge signatures. RSA and DSA/ECDSA are commonly used signature schemes.

\subsection{Two-Factor Authentication}
\label{sec:2fa}

Broadly defined, two-factor authentication (2FA) refers to an authentication protocol that requires 
a user to login to a service by proving knowledge of a 
password (first factor, ``something you know'') and possession of a previously registered personal device (second factor, ``something you have'').



\mypara{\stfa Protocol}
\stfa\ is the most prominent two-factor authentication in which the user's personal device generates a one-time PIN. The PIN is transferred to the server for verification, often manually with the user's assistance and through the login terminal. To verify the device, the server generates the same PIN and compares it with the one it receives from the terminal\footnote{The deprecated SMS 2FA \cite{nist-sms-2fa} uses a similar protocol flow, but the device does not compute the PIN. 
Rather, the PIN is communicated by the server and sent to the user's registered device as a text message.}.


A popular PIN generation function is HMAC \cite{hmac-rfc}. In an initial setup phase, the server and the device agree on an HMAC key. TOTP \cite{totp} defines the timestamp and HOTP \cite{hotp} defines a counter as the HMAC message. In general, it could be any parameter known both by the server and the device.
The HMAC output is encoded into a number and truncated to a small number of digits (e.g., six digits) for easier manual copying by the user. 
\mypara{\ntfa Protocol}
\ntfa\ is a relatively new trend in the 2FA design-space, 
which attempts to address the usability issues with \stfa. 
This method 
follows a simple request/response message exchange between the server and the device. 
Upon receiving a login request, 
the server sends a notification message to the registered secondary device, and 
waits for the user to approve the validity of the login attempt. It is assumed that the user only approves  
 login requests initiated by herself and denies any other attempt. The approval (or denial) response 
 proves to the server that the user possesses the secondary device. 
 Since the user interaction with the system is limited to approving or denying a request, 
 displayed as a notification on the user's device, \ntfa\ is considered more user-friendly than \stfa \cite{duopushusability, abbott2020mandatory}.
 
\ntfa\ assumes that the server has a direct or indirect connection with the device. \shashank{We need this too. \ntfa needs more.}
This connection is commonly supported by third-party services such as 2FA service providers
(e.g., Authy \cite{twilio-push}, Google \cite{googleprompt}, Duo \cite{duo-push}) and push notification services (e.g., Google cloud messaging \cite{google-cloud-messaging}, Apple notification \cite{apple-notifications}). Without such third party service, the server 
needs to establish a connection directly with the device. Establishing such a connection is not a trivial engineering effort
since smartphones (assuming that 2FA system uses a smartphone) 
 do not keep a permanent internet address and migrate from one network to another constantly.

\mypara{FIDO U2F}
2FA protocols could also be based on asymmetric-key authentication schemes such as digital signatures. In FIDO U2F \cite{u2f-tech-overview}, for example, the device generates a public/private key pair during registration and transfers the public key to the server. During authentication, 
the server sends a challenge to the user's device through the client. The user locally authenticates to her trusted device to sign the challenge and transfers the response to the server through the client. This approach requires support from the client browser and establishment of a secure device-client channel.

\subsection{Benefits and Limitations of Current 2FAs}
\label{sec:related}

\mypara{\stfa} 2FA hardware tokens such as RSA Secure ID \cite{sid}, and mobile apps, such as Google Authenticator \cite{google-authenticator} and HID ActiveID \cite{hid},
generate a one-time short PIN that is manually copied by the user to the login terminal. 
The device generating the PIN only needs to be capable of generating the PIN (not transferring it) and 
be equipped with a display to show it. Therefore, this type of 2FA systems are suitable when  
offline devices or devices with limited connectivity are used as second factor.

The main user interaction in \stfa\ is manual copy of PINs, which limits their length, increasing the risk of guessing attacks.
In contrast, \ptfa\ does not require manual entry of PINs, and therefore, PINs could be long to offer much better 
resistance to guessing attacks. Shoulder surfing of PINs is another issue 
with \stfa \cite{Eiband:2017:USS:3025453.3025636}, which does not arise in \ptfa.

\mypara{\ntfa} In recent years, smartphones with more connectivity options 
created opportunities for more usable approaches.
2FA providers, such as Authy \cite{authy} and Duo \cite{duo}, offer several options 
for simpler user interaction.  Push-based 2FA offered by these providers lets users approve or 
deny a 2FA request notification sent to their phone. While user engagement is much simpler compared to traditional PIN-based 2FA, a neglectful user could mistakenly accept an attacked session since the binding between 
the $\Cl$-$\Sr$ login session and $\Dv$-$\Sr$ approval, relies only on user's attention. 

To maintain a connection from the server to the device, a secure and trusted channel should be established, imposing unwanted trust 
on the apps and push service providers, which are shown to be target of various attacks \cite{li2014mayhem, loreti2018push,xu2012abusing, pushattack1, pushattack2}. These channels may also be unreliable as shown in \cite{abbott2020mandatory}
 who reported 20\% failed authentication due to ``no response'', caused by the users not responding in time (due to errors in notification communication). 
In contrast, \ptfa\ establishes a strong binding between the $\Cl$-$\Sr$ login session and $\Dv$-$\Sr$ communication 
by requiring users to input the displayed \id, 
which avoid attacks arising from users neglectfully accepting attacker's sessions \cite{push-accept-dark-side}. 
Besides, messages are communicated from $\Dv$ to $\Sr$ directly with no security requirement 
on the channel or reliance on third party services. 

\mypara{FIDO U2F} Other approaches have been introduced that establish a channel between the device and client to automatically 
transfer some token through the client to the server on user's approval. In FIDO, for example, a channel is established between user's trusted device (e.g., smartphone, security key) and the client \cite{u2f-tech-overview}. 
Other challenge-response 2FA mechanisms \cite{czeskis2012strengthening,ndss-2014} also rely on 
establishment of a channel between the device and client to transfer the response. 
In \ptfa, $\Dv$'s response to $\Sr$ is not channeled through the client. Therefore, \ptfa does not require establishing a secure $\Dv$ to $\Cl$ channel. Thus, it makes migration to new clients hassle-free.  

\hit{On the flip side, FIDO U2F is resistant to phishing but \ptfa isn't, much like \stfa. FIDO U2F binds the user login to the origin  since the client sends the origin and the channel ID as part of the challenge to the device and the device signs the entire challenge.}


\section{\ptfa\ System \& Design Goals}
\label{sec:model}


\subsection{System Model}
\label{sec:sysmodel}


There are four parties involved in \ptfa\ to securely authenticate a user to a service, 
as shown in Figure \ref{fig:system}. 
A ``user'' $\Us$  initiates a request from a ``client'' $\Cl$ 
(a.k.a.\ the client terminal, like a web browser on a laptop) by entering her username and password, 
as is the case with any password authentication protocol. $\Cl$ submits the request 
to a ``server'' $\Sr$ (a.k.a.\ a service provider, hosted on a web server for example) over a secure 
channel. $\Us$ also proves to $\Sr$ the possession of a secondary factor personal device 
$\Dv$ (e.g., smartphone, smartwatch) by entering a \id on $\Dv$, which is provided to $\Us$ by  $\Sr$ through $\Cl$.

In our model, the client and the server interact over a bidirectional $\Cl \Leftrightarrow \Sr$ channel. 
As in other password authentication protocols, we assume a secure $\Cl \Leftrightarrow \Sr$ channel
such as a TLS channel. $\Cl$ and $\Dv$ interaction is assisted by the human user 
(i.e., a manual $\Cl  \rightarrow \Dv$ channel), who observes the \id displayed on the client 
and inputs the same on the device. 
Finally, the device communicates to the server to 
manifest the knowledge of the correct \id and PIN  through $\Dv \rightarrow \Sr$ channel, e.g.,  internet. 

Other 2FA systems usually involve a similar set of parties but the communication channels may be 
different.
For example, in the traditional \stfa, the $\Dv  \Rightarrow  \Cl$ secure communication
is commonly assisted by a human user and happens in one direction, from the device to the client (i.e., the user copies a 
PIN from the device to the client terminal). In contrast, in \ptfa\ the manual unprotected unidirectional $\Cl  \rightarrow  \Dv$ channel 
is for transferring the \id from the client to the device. \ntfa\ relies on third party push notification 
service providers (e.g., Firebase Cloud Messaging \cite{google-cloud-messaging}) to handle secure messaging between the device and the server. In our 
model, an ordinary channel from device to server is good enough, which could be established without relying on third-party services.


\subsection{System Overview}
\label{sec:overview}

Our system consists of two phases (a detailed discussion can be found in Section \ref{sec:system}):

\mypara{Registration} The initial \textit{``registration phase''} through which the parties in the protocol establish the communication channels and share
the 2FA secret keys. 
\shashank{I think sharing cryptographic keys implicitly establishes secure channels.} 
During this phase, the user registers a username and a password with the server, 
as the first authentication factor. The server and the device agree on a secret key (typically picked by the server and transferred to the device manually  or by scanning a QR Code)\shashank{by the user?}, which will be used later to generate one-time PINs. 
The server stores the user's state (e.g., hash of the password and the secret key) and the device stores the secret key associated with the user account on the server. 
The client does not store any state during this phase. 

\mypara{Authentication} The \textit{``authentication phase''}  through which the parties interact to securely authenticate the user to the server. 
During this phase, the registered username and password are used as the first factor, and the server displays 
a unique
\id to the user through the client machine (e.g., on the login webpage). The user \hit{specifies the server name on the device} and inputs the \id, 
 which consequently sends the \id and a PIN to the server to  
 prove the presence of the device. 

Entering the \id indicates that the user has diligently approved the submission of the PIN. 
Besides, the uniqueness of the \id helps the server to distinguish between two concurrent sessions and 
binds the received PIN to the associated session.
Upon receiving the authentication information, the server  verifies the password and the PIN, and authenticates the user's  session associated with the \id. 

During the authentication process, the server keeps a temporary record of all active sessions and \ids.

\subsection{Design Goals}
\label{sec:design-goals}

The overall goal of the system is to securely authenticate a user to a service 
using  two authentication factors that are  known (i.e., password) and owned by the user (i.e., a device generating a one-time PIN). 
An attacker with access to only one of the two factors cannot authenticate to the system without launching an 
online guessing attack on the other.
From a usability perspective, we expect the system to offer a high level of usability (at least comparable to that of traditional \stfa or perhaps higher since the \id unlike PIN is not a secret, and hence can be short and simple). 

\subsubsection{Security Goals}
We consider various ways in which the system could be attacked: any of the entities could be compromised or the channels connecting them. We model the attacks with the help of a (computationally-bounded) attacker $\adv$. The primary goal of $\adv$ is to get access to a user $\user$'s account.  


\mypara{Client compromise} An attacker $\adv$ compromises a client from which $\user$ logs into $\Sr$, and learns $\user$'s password. Even with this knowledge, it should be \emph{infeasible} for $\adv$ to get access to $\user$'s account (assuming the registered device is not compromised). 

\quad $\adv$ could also try to log in at the same time as $\user$. In other words, it may be able to run authentication sessions with the server concurrently with the user. Even when it can do so, it should remain infeasible to get into $\user$'s account. (Here, we will assume that $\user$ does not make any mistake entering the displayed \id on her device.)


\mypara{Device compromise} An attacker $\adv$ compromises a user $\user$'s device $\Dv$, which enables it to control the device fully. In particular, $\adv$ is able to authenticate to $\Dv$ (if needed) and get access to any secret information stored on the device (directly or otherwise). We want that $\adv$ does not gain any knowledge of $\user$'s password as a result of this. Specifically, $\adv$ could only attack $\user$ by guessing her password in an online manner.

\mypara{Channel compromise} Even when none of the  entities in the system (client, device, or server) are under attack, an adversary $\adv$ could have control over the channels connecting them. It may just eavesdrop on the channels passively, or could modify/censor the traffic actively. Irrespective of how powerful $\adv$ is, we want to make sure that it does not gain any advantage over attackers who have no access to the channels.



\mypara{Attacks on third parties} As we have seen several times, there are three main entities in a 2FA system: client, device and server. Several 2FA schemes, however, introduce additional entities like a 2FA service provider. We can either trust these other entities or study how the security of the system is affected when they are targeted by attackers. The attacks could be targeted at the entities themselves or the channels connecting them with others. As a result, analyzing the security of the system becomes substantially more complex. Therefore, we want to design a system that does not introduce any other entities beyond the three main ones. Such a system would keep the attack surface low.

\mypara{User negligence} Users can be negligent sometimes, granting access to others unintentionally. We would like the 2FA scheme to make such 
oversights
harder for users.


\begin{figure*}
\boxed{
\small
\begin{tabular}{p{\textwidth}}

\noindent\textbf{Parties.}
The protocol runs between a user (who knows a password and owns a device), and a server, through a client machine.  The four parties involved in the protocol are: 

\begin{itemize}
\item Server $\Sr$ offering a service to authorized registered users. 
\item User $\Us$ who attempts to register an account with $\Sr$ or seek an authorized access to the service offered by $\Sr$.
\item Client $\Cl$ using which $\Us$ submits a registration or access request to $\Sr$. 
\item Device $\Dv$ which is a user personal 
device (e.g., smartphone) used in the authentication process.
\end{itemize}

\rule{\textwidth}{0.5pt}

\noindent\textbf{Parameters.}
The 2FA protocol is based on  well-known password authentication and   
one-time PIN authentication mechanisms and has the following parameters.

\begin{itemize}
\item Password $\pw$ from a dictionary $\mathsf{dict}_{\pw}$ of size $d$.
\item Random secret 2FA key $k$ of size $m$ from $\{0,1\}^m$.
\item One-time password generator function $F_k(x)$ computed as HMAC$(k, x)$, where $x$ is a time reference for TOTP or a counter reference for HOTP algorithm. Using this function, a one-time $\pn$ is computed as $\pn = F_k(x|\pt)$, and optionally truncated to $n$-bits. 
\item A temporary \id $\pt$ chosen from a domain $\mathsf{dict}_{\pt}$ of size $p$.
\end{itemize}

\rule{\textwidth}{0.5pt}

\noindent\textbf{Registration. } The registration protocol runs between the parties to register the user known by a username $\uname$ and a password $\pw$ 
with the service $\Sr$ and to share the secret $k$ between $\Dv$ and $\Sr$. 
One round of authentication protocol (detailed next) should be completed to ensure the success of the registration.  

\rule{\textwidth}{0.5pt}

\noindent\textbf{Authentication. }
The authentication protocol runs between the four parties to 
authenticate the user with the help of the shared secret parameters according to these steps:


\noindent \textit{Step 1.} On the Client 
\begin{itemize}
\item $\Cl$ establishes a secure authenticated channel with $\Sr$. \hit{Let $\sname$ be the common name for $\Sr$.}
\item $\Us$ enters $\uname$ and $\pw$ on $\Cl$. 
\item $\Cl$ submits $\uname$ and $\pw$ to $\Sr$. 
\end{itemize}
\noindent \textit{Step 2.} On the Server 
\begin{itemize}
\item $\Sr$ verifies $\pw$.
\item $\Sr$ starts an authentication session timer $t_{s}$. It picks a fresh $\pt$ from $\mathsf{dict}_{\pt}$ and sends it to $\Cl$, which is  displayed to $\Us$. 
\end{itemize}
\noindent \textit{Step 3.} On the Device 
\begin{itemize}
\item $\Us$ specifies $\uname$ and $\sname$ on $\Dv$ and inputs the displayed $\pt$.
\item $\Dv$ computes $\pn$.
\item $\Dv$ 
submits $\pn$ and $\pt$ to $\Sr$.
\end{itemize}
\noindent \textit{Step 4.} On the Server
\begin{itemize}
\item $\Sr$ verifies $\pn$ (server may delay verification of $\pw$ till this point). \sha{combine with below? separately, doesn't make much sense}
\item $\Sr$ accepts the authentication session associated with $\pt$ if $\pn$ and $\pw$ are correct, and rejects the session otherwise. 
 $\Sr$ starts a timer to discard temporarily stored $\pt$ on successful/failed/expired authentication.  \shashank{Server should also use $\pt$ in this step. Should we say something like server uses \id to locate the right session to authenticate.}
 \end{itemize}

\end{tabular}}
\caption{\ptfa\ registration and authentication protocols.}
\label{fig:prot}
\end{figure*}

\subsubsection{Usability and Deployability Goals} 
We attempt to make the system easy to be used and to be deployed:

%

\mypara{Ease of use} After the initial setup, the system should be effortless for frequent use. Interaction of users with the system should be maintained at the minimum level required to achieve the security goals. 
A user with average technical background should be confident using the system if sufficient instructions are provided.

\mypara{Easy integration} Client-side modification should not be required.
Server-side modification should be minimal.
Similar to traditional \stfa, the server should just need to verify a password and  
a one-time PIN. 

\mypara{Universal compatibility} Users should be able to register devices with different input modes (touchscreen, camera, keypad, etc.) as the second factor.




\section{\ptfa Protocol}
\label{sec:system}

\begin{figure}[t]
\centering
\includegraphics[width=0.85\textwidth]{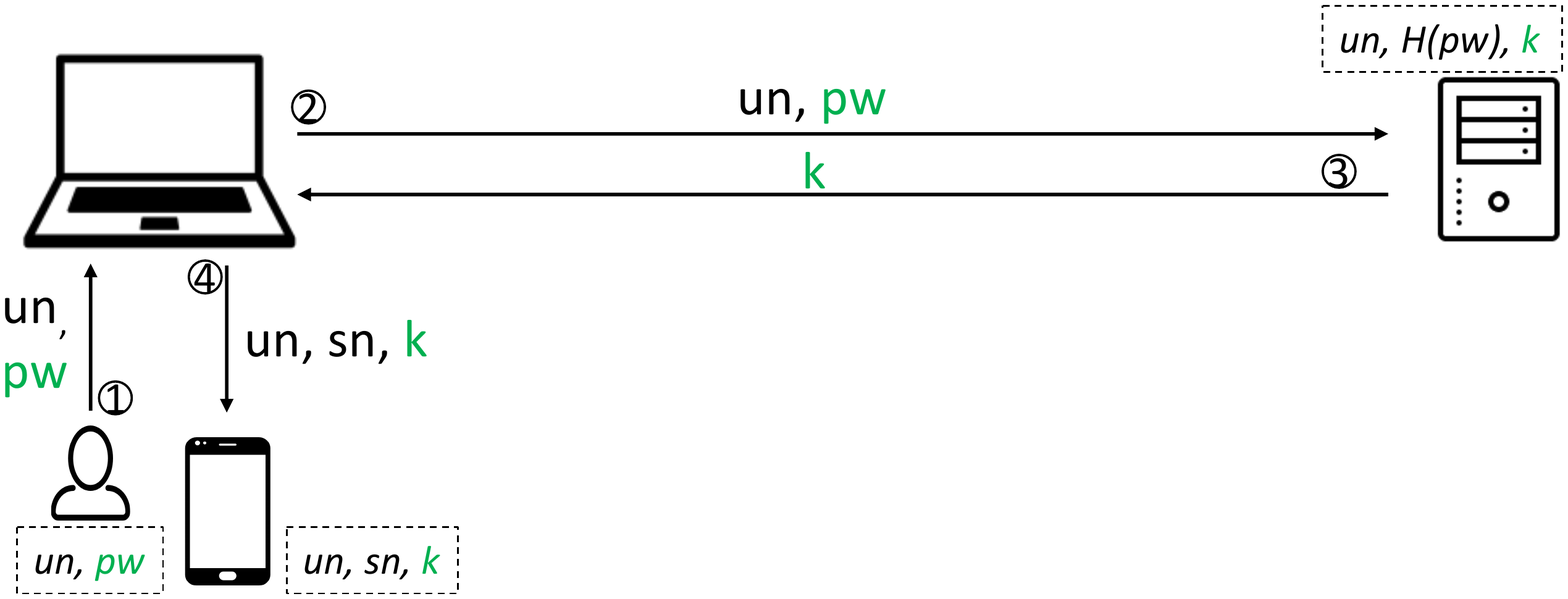}
\vspace{-36mm}
\caption{\ptfa\ registration overview.}
\label{fig:register}
\end{figure}

\begin{figure}[t]
\centering
\includegraphics[width=0.85\textwidth]{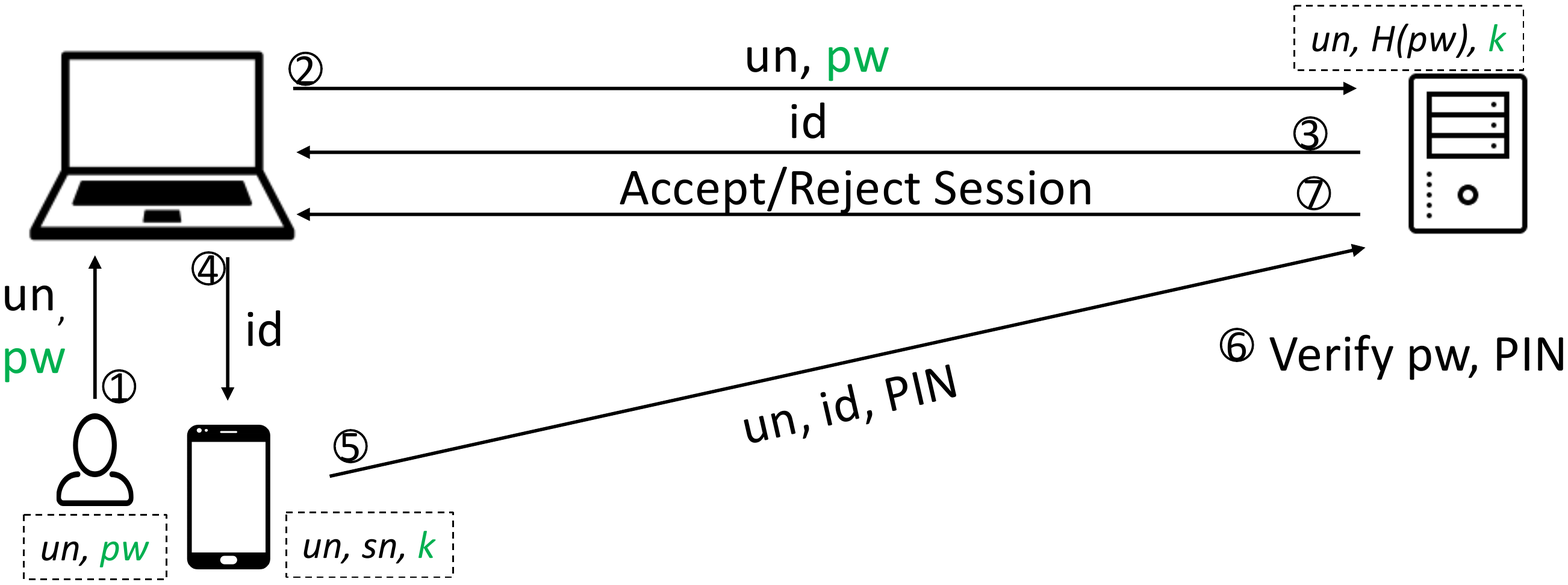}
\vspace{-36mm}
\caption{\ptfa\ authentication overview.}
\label{fig:auth}
\end{figure}

In this section, we present the protocol description in Figure \ref{fig:prot} and 
 elaborate on the registration and authentication phases in Section \ref{sec:sys-reg} and \ref{sec:sys-auth}, respectively.




\subsection{Registration Protocol} 
\label{sec:sys-reg}

A high level overview of the registration flow is shown in Figure \ref{fig:register}. 
We provide some insight on the 
choices we made in designing the protocol to achieve our goals.  

We assume that a user $\Us$ initiates the registration process from only one client at a time 
(no concurrent registration).\footnote{
\hit{
The server can simply avoid concurrent registrations since only one registration per account is generally needed. 
However, the user may want to login to the same service from multiple clients (e.g., sharing files between two machines).
 Therefore, we do consider concurrent logins.}}
During the process, she communicates with a server $\Sr$ to register an account known by 
a username and authenticated by a password. Most of the services deploying two-factor authentication separate password registration from secondary device registration. In our protocol, we assume that the user registers 
the password and the secondary device during the same setup phase.  However, as with other 2FAs, the user may enroll into two-factor authentication at any point after registering for password-only authentication.

The goal of the registration process is to store user's state with the server, and share the secrets required to 
authenticate the user. Therefore, it is important to establish a secure communication channel. Since the registration is a one-time task, we can assume that  all efforts are made to make the transfer of the secrets secure. 
For example, we may assume higher engagement from the user since the overall 
impact on user experience is negligible.

The transfer of the 2FA secret key from the server to the device can happen through the client and with the assistance of the user. 
For example, the client 
can display the secret received from the server (encoded into plain text or QR Code for example) 
 and ask the user to input it  on the device. 
 \shashank{Is this how it usually happens or are we suggesting that this is how it should happen? Also, may be we shouldn't use camera for this because that's one of the things we want to avoid.}
\hit{In a similar way, the user can be asked to save some information about the server  (e.g., the server domain name -- also known as $\sname$ in our protocol)  and 
herself (e.g., username, or account nickname -- also known as $\uname$ in our protocol).
This information is essential to distinguish between different services on the same personal device. } 

Once the 2FA secret key is transferred, one round of 2FA may be carried out to ensure the accurate transfer of information. A complete round of 
\ptfa\ consists of picking a unique \id by the server, submitted to and displayed on the client for the user to transfer it to the device (e.g., by drawing a pattern on the device touch-screen). Once the device receives the  \id for a specific service, it computes a one-time PIN, 
and submits the PIN and the \id to 
the server. If the PIN is correct, the server accepts the connection associated with the \id.

\shashank{Should we discuss that in the registration phase, some information would be 
stored on the device which would help it to establish a secure channel to the server over the internet? What kind of protocols are used for this?}
\maliheh{lets discuss this communication channel}


\subsection{Authentication Protocol}
\label{sec:sys-auth}

The high level authentication flow is shown in Figure \ref{fig:auth} and the protocol is described in Figure \ref{fig:prot}. To avoid repetition, we do not discuss the protocol details here. Instead, we focus on the design choices.


To authenticate successfully to a service, a user inputs a username and password on a client, and copies an \id displayed on the client to a registered device.
The PIN is computed on the device and is transferred to the server automatically (unlike traditional \stfa\ where the user inputs the PIN manually). Receiving the correct PIN proves that the user holds the device and has explicitly consented to the authentication process.
 
In \ntfa\ systems, a similar user consent is provided by asking the user to approve or deny a message sent to her device. We believe that the user interaction in \ptfa is less prone to user errors compared to tapping an approve/deny 
button. Intuitively, the possibility of a neglectful user mistakenly accepting an attacker's session by approving a notification message sent to the device for the attacker's login attempt is 
higher than the user deliberately inputting an \id that matches the one generated for the attacker's session. 
 
\begin{figure}[t]
\centering
\includegraphics[width=0.85\textwidth]{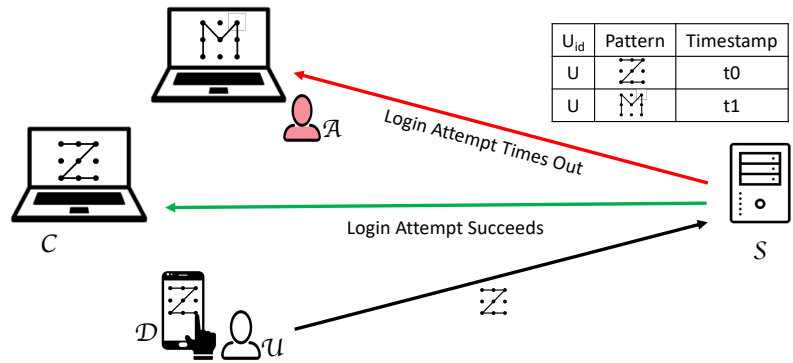}
\vspace{-30mm}
\caption{Role of \ids in distinguishing concurrent sessions. We take the example of drawing patterns here.}
\label{fig:concurrent}
\end{figure}

Therefore, besides serving as a proof of consent, \ids distinguish between concurrent sessions belonging to one user. 
Let us consider a user who tries to login to her account from multiple devices concurrently, for example, a smartphone and 
a laptop. Once the server receives an \id, it compares the \id with the list of \ids that have been generated 
and displayed to the user, and accepts only the session associated with the received \id. 
As an example, if the displayed \id is the pattern $\mathsf{Z}$ for the smartphone and $\mathsf{M}$ for the laptop, drawing 
 $\mathsf{Z}$ only approves the smartphone session (see Figure \ref{fig:concurrent}). Now, if one of the concurrent sessions belongs to an attacker 
who is planning his/her login to occur concurrently with the user, it would be unlikely that the user draws
 the pattern belonging to the attacker's session.
 \hit{Note that the uniqueness of identifiers applies to concurrent sessions belonging to the \emph{same} user ($\uname$).  The same set of identifiers could be used for all users registered with a service and the same identifier might be issued 
simultaneously to multiple users. }

 
In-use identifiers should not only be ``unique'' among all concurrent sessions, 
but also distinct enough with an optimal distance according to a ``similarity metric''\footnote{In Appendix \ref{sec:patternchoice}, 
we define the ``similarity metric'' between two patterns according to the number of slips it takes to convert one pattern to another. 
A lower number of slips corresponds to a higher similarity metric. }.
The distance is important to minimize manual input errors that may lead to acceptance of an attacker's session. 
For example, $\mathsf{Z}$ and $\mathsf{7}$
are more similar compared to $\mathsf{Z}$ and $\mathsf{M}$, so the former has more room for user errors. 
Therefore, the server 
should not issue  $\mathsf{Z}$ and $\mathsf{7}$ for two concurrent sessions. 
\hit{This consideration is only valid if identifiers are input manually, for example by drawing a pattern or entering a numerical value. 
Human input errors are not relevant if more automated tools such as QR Codes scanning is used.}

The ``optimal distance'' and ``identifier uniqueness' limit the number of identifiers that can be in-use. 
In addition to these two constraints, we should be mindful of the usability of manually inputting the identifier and 
restrict the identifiers to those that are easy to input (for example by restricting the identifier length), which could 
further limit the number of possible identifiers. 
\nsha{Previous sentence may need some fixing.} 

The communicated identifiers would of course not immediately be ready for future use, 
to avoid replay attacks as we explain next.   
2FA systems usually allow the server to accept one or more PINs prior and after the current PIN
to compensate for time drifts between the device and server. An attacker may replay a
 PIN and identifier within the time window considered for the time drift and succeed. 
Hence, we start a timer after completion or expiration of an authentication session. The timer 
runs for a duration  greater than an acceptable time drift window. 
\nsha{Why $>$ here?}
The communicated 
identifier will be ready for reuse once the timer matures.  For example, assuming an 
allowable time drift of 60s backward and 60s forward, 
the identifier will be ready for reuse 120s after the completion/expiration 
 of an authentication session that used the identifier.


\hit{
A question that may arise is whether limiting the number of possible identifiers 
may give a DoS attacker the opportunity to open several concurrent sessions to 
mark all possible identifiers as in-use, and therefore, stop a legitimate user from logging in. 
Note that since typically the number of allowed concurrent logins for one user are not boundless 
\hit{(to avoid DoS and password guessing attacks)}, a limited number of in-use identifiers should not be problematic. 
The identifier domain size can be defined by the number of concurrent sessions that a service provider allows per each user. 
For example, for a server that
allows only 100 simultaneous sessions for a user, a set of 100 \ids should be sufficient since the 101$^{\text{st}}$ session   
is rejected by the service provider anyway (not due to lack of available identifiers but to avoid password guessing and DoS attacks). 
  }


\section{Security Analysis}
\label{sec:security-analysis}

We analyze the security of \ptfa\ against the attacks  described in Section \ref{sec:design-goals}.

\mypara{Client compromise.} With $\user$'s password, $\adv$ will clear the first hurdle of authentication with $\Sr$, and receive an \id $\pt$ in return. To complete the authentication process, $\adv$ needs to provide a valid PIN corresponding to some `fresh value' (a new value of time-stamp, new counter value, etc.). However, without knowledge of HMAC key, $\adv$ has \emph{negligible} chance of doing so. 

If $\adv$ can log in concurrently with $\user$, then the `fresh value' could be the same for both (e.g., when TOTP is used). So a PIN sent from user device $\Dv$ could also be a valid PIN for $\adv$'s session. However, there are two important things to note here. First, apart from the `fresh value', \id is also involved in PIN computation. Second, if $\adv$ logs in first (from $\Sr$'s perspective) and its session has not expired, a different \id is sent to $\user$ (or, vice versa, if $\user$'s logs in first and has an active session, a different \id is sent to $\adv$). Therefore, even when $\adv$ and $\user$ log in at about the same time, they receive different \ids. As a result, the PIN sent by $\user$ has negligible chance of being a valid PIN for $\adv$'s session. 

\mypara{Device compromise.} When the registered device $\Dv$ is compromised, $\adv$ can get the hold of the HMAC secret key. Hence, it can generate PIN on any value of its choice. However, no information about $\user$'s password is ever stored on $\Dv$\textemdash not even temporarily. So $\adv$ can only try to guess the password in an online manner.

\mypara{Channel compromise.} Three different channels are used in the design of \ptfa: between $\Cl$ and $\Sr$, between $\Cl$ and $\Dv$, and between $\Dv$ and $\Sr$. The first one is a secure channel, which cannot be compromised. The second one is a regular channel through $\user$. $\adv$ can eavesdrop on $\user$ to observe the \ids she inputs. The channel from $\Dv$ to $\Sr$ is also a regular channel.  
$\adv$ can gather (\id, PIN) pairs from the channel, and use them later in any manner it pleases. To make the attack more severe, let us also assume that $\adv$ knows the first factor. (There is no point assuming that $\adv$ knows the HMAC key since spying on the channel would be meaningless.)

The analysis here is not very different from the first case above (client compromise). $\adv$ must generate a valid PIN on a `fresh value' to successfully authenticate. This value would be different from the values corresponding to the PINs collected from the channel, unless those PINs were generated in about the same time frame as the attacker's session and the values used to generate PINs are time-stamps (like in TOTP). When this is true, we can again rely on the fact that no two unexpired authentication sessions share the same \id. Thus, even when a PIN on the channel corresponds to the same time-stamp that $\adv$ needs, it will correspond to a different \id, making the task of PIN scraping on the device-server channel unfruitful.

\mypara{User negligence.} Users can sometimes accidentally approve an attacker's session. In the case of \ptfa, this could only happen when a user inputs an \id that does not match with the one displayed to her, but it matches with the one given to the attacker. Depending on the type of \id used, the frequency of such errors could be very small. \nsha{Cite the study here}


\mypara{Attacks on third parties.} A central benefit of \ptfa is that it does not introduce new entities into the authentication infrastructure beyond the ones that are already well-established. In other 2FA schemes, one has to rely on one or more intermediaries between device and server to secure the communication between them. \shashank{give examples, say what exactly they do} In our case, however, no special security property is needed for this channel. We just need the channel to deliver messages to the server.








\subsection{Comparative Analysis} We use the framework of Bonneau et al.\ \cite{bonneau2012quest} to compare \ptfa with Sound-Proof \cite{karapanos2015sound}, FBD-WF-WF \cite{shirvanian2014two}, FIDO U2F \cite{u2f-tech-overview}, \stfa, and \ntfa. The framework defines 25 properties framed as a diverse set of benefits and a methodology for comparative evaluation. We have borrowed from similar analyses done in prior work, and from the Sound-Proof paper \cite{karapanos2015sound} in particular, but there are some differences. 


A \stfa scheme can be implemented in various ways. PINs can be delivered via SMS or voice, or generated on the device itself. Here, we consider PIN generation on the device through an app (like Google Authenticator \cite{google-authenticator}) since SMS 2FA is deprecated \cite{nist-sms-2fa}. For FIDO U2F, we assume  a smart phone as the authenticator although other form factors such as security tokens are also possible. 

In Table \ref{comparative-analysis}, for each 2FA method, we state whether the method offers the benefit or somewhat offers the benefit or does not offer the benefit at all. Here we summarize the results.



\newcommand{\cS}{$\circletfillhl$}
\newcommand{\cY}{$\circletfill$}
\begin{table}
\tiny
\caption{Comparison of \ptfa against Sound-Proof \cite{karapanos2015sound}, FBD-WF-WF \cite{shirvanian2014two}, FIDO U2F \cite{u2f-tech-overview}, \stfa, and \ntfa using the framework of Bonneau et al.\ \cite{bonneau2012quest}. We use \cY{} to denote that method offers the benefit and \cS{} to denote that the method somewhat offers the benefit.}
\label{comparative-analysis}
  \centering
\begin{tabularx}{\textwidth}{lXXXXXXXXcXXXXXXcXXXXXXXXXXX}
\toprule
&\multicolumn{8}{c}{\textbf{Usability}}&\multirow{8}{*}{~}&
    \multicolumn{6}{c}{\textbf{Deployability}}&\multirow{8}{*}{~}&\multicolumn{11}{c}{\textbf{Security}}
    \\
 \cmidrule{2-9} \cmidrule{11-16} \cmidrule{18-28}
 \textbf{Scheme}& 
    \rotatebox[origin=l]{90}{Memorywise-Effortless} & 
    \rotatebox[origin=l]{90}{Scalable-for-Users}  & 
    \rotatebox[origin=l]{90}{Nothing-to-Carry}&
    \rotatebox[origin=l]{90}{Physically-Effortless} & 
    \rotatebox[origin=l]{90}{Easy-to-Learn}& 
    \rotatebox[origin=l]{90}{Efficient-to-Use}& 
    \rotatebox[origin=l]{90}{Infrequent-Errors}& 
    \rotatebox[origin=l]{90}{Easy-Recovery-from-Loss} &
    & 
    \rotatebox[origin=l]{90}{Accessible}&
    \rotatebox[origin=l]{90}{Negligible-Cost-per-User}&
    \rotatebox[origin=l]{90}{Server-Compatible}&
    \rotatebox[origin=l]{90}{Browser-Compatible}&
    \rotatebox[origin=l]{90}{Mature}&
    \rotatebox[origin=l]{90}{Non-Proprietary}&
    & 
    \rotatebox[origin=l]{90}{Resilient-to-Physical-Observation}&
    \rotatebox[origin=l]{90}{Resilient-to-Targeted-Impersonation}&
    \rotatebox[origin=l]{90}{Resilient-to-Throttled-Guessing}&
    \rotatebox[origin=l]{90}{Resilient-to-Unthrottled-Guessing}&
    \rotatebox[origin=l]{90}{Resilient-to-Internal-Observation}&
    \rotatebox[origin=l]{90}{Resilient-to-Leaks-from-Other-Verifiers}&
    \rotatebox[origin=l]{90}{Resilient-to-Phishing}&
    \rotatebox[origin=l]{90}{Resilient-to-Theft}&
    \rotatebox[origin=l]{90}{No-Trusted-Third-Party}&
    \rotatebox[origin=l]{90}{Requiring-Explicit-Consent}&
    \rotatebox[origin=l]{90}{Unlinkable}
    \\
 \midrule
 Sound-Proof	&   &   &\cS&   &\cY&\cY&\cS&\cS&   &\cY&\cY&   &\cY&   &\cY&   &   &\cS&\cY&\cY&   &\cY&   &\cY&\cY&\cY&\cY\\
 FBD-WF-WF 		&   &   &\cS&   &\cY&\cY&\cS&\cS&   &\cY&\cY&   &   &   &\cY&   &   &\cS&\cY&\cY&   &\cY&   &\cY&\cY&\cY&\cY\\
 FIDO U2F     	&   &   &\cS&   &\cY&\cY&\cS&\cS&   &\cY&\cY&   &   &\cY&\cY&   &\cY&\cS&\cY&\cY&   &\cY&\cY&\cY&\cY&\cY&\cY\\
 \stfa    		&   &   &\cS&   &\cY&\cS&\cS&\cS&   &\cS&\cY&   &\cY&\cY&\cY&   &   &\cS&\cY&\cY&   &\cY&   &\cY&\cY&\cY&\cY\\
 \ntfa       	&   &   &\cS&   &\cY&\cY&\cS&\cS&   &\cY&\cY&   &\cY&\cY&\cY&   &\cY&\cS&\cY&\cY&\cY&\cY&   &\cY&   &\cY&\cY\\
 \ptfa         	&   &   &\cS&   &\cY&\cY&\cS&\cS&   &\cY&\cY&   &\cY&   &\cY&   &\cY&\cS&\cY&\cY&\cY&\cY&   &\cY&\cY&\cY&\cY\\
\bottomrule
\end{tabularx}
\end{table}

\mypara{Usability.} Several usability benefits are offered at the same level by all the methods. None of the methods are \emph{Memorywise-Effortless}, \emph{Scalable-for-Users}, or \emph{Physically-Effortless} because they do not replace passwords. The benefit \emph{Nothing-to-Carry} is somewhat offered because a common device like smartphone can be used as the second factor. In terms of being \emph{Easy-to-Learn}, \emph{Infrequent-Errors}, and \emph{Easy-Recovery-from-Loss}, all the methods are roughly the same. While most methods are \emph{Efficient-to-Use}, \stfa is only somewhat efficient because one has to copy PINs manually. With \ptfa, one can use QR codes, patterns, etc.\ to keep the authentication time short. 

\mypara{Deployability.} Except \stfa, all the methods are \emph{Accessible} because users have to do little beyond typing passwords. The cost per user for all the methods is negligible because users carry a secondary factor like smartphones anyway. None of the methods are server compatible as the provider needs to handle something other than passwords. Apart from FIDO U2F and FBD-WF-WF, none of the methods require any changes to the client, so they are browser compatible. In terms of maturity, FIDO U2F, \stfa, and \ntfa are very mature technologies. All the methods are non-proprietary.


\mypara{Security.} To be \emph{Resilient-to-Physical-Observation}, an attacker should not be able to impersonate a user after observing them authenticate one or more times. Various types of attacks are included here like shoulder surfing, filming the keyboard, recording keystroke sounds, etc. 
Sound-Proof does not provide this benefit because it suffers from co-location attacks. FBD-WF-WF and \stfa do not provide it either because they are vulnerable to shoulder-surfing attacks. To be \emph{Resilient-to-Targeted-Impersonation}, an acquaintance should not be able to impersonate a specific user by exploiting knowledge of personal details like date of birth. All the methods provide this benefit somewhat only because victim's personal details could be used for (fake) account recovery.

Most of the schemes do not provide the benefit \emph{Resilient-to-Internal-Observation} because PIN is transferred through the client machine. On the contrary, in both \ntfa and \ptfa, users do not input anything beyond passwords on the client. In \ntfa, users approve or reject notifications on the registered device and, in \ptfa, they input \ids on the device. 

Apart from FIDO, none of the methods are resilient to phishing attacks. FIDO provides security against phishing by including the origin address in the signed tokens. While most of the methods do not rely on trusted third parties, \ntfa usually does. All the methods are unlinkable because the authenticators (PINs, identifiers, etc.) cannot be linked across multiple providers.


\section{System Implementation}
\label{sec:deploy}


As a proof of concept, we designed and developed a web-based authentication service. We developed the server side scripts running on a Macbook Pro Mojave 10.14.5 with 2.9 Intel Core i7 processor and 16 GB 2133 MHz memory. The 2FA application was developed in Java for Android on a Huawei Honor 7X smartphone with Kirin 659 processor and 3GB memory.


\subsection{Choice of Identifier}
\label{sec:idchoice}

\ptfa\ relies on the user to copy the \id displayed on the client to their device to \emph{confirm} 
 the association of the client-server authentication session with the device-server session. 
 \Id values could have a short textual or graphical representation (e.g., few-digit number, letters, a short phrase). 
To reduce the users' cognitive load we suggest using common input methods. In our implementation, we considered patterns and QR Codes, 
and from now on we refer to them as PT-2D-2FA and QR-2D-2FA respectively. 



\begin{figure}[t]
\centering
\includegraphics[width=0.95\columnwidth]{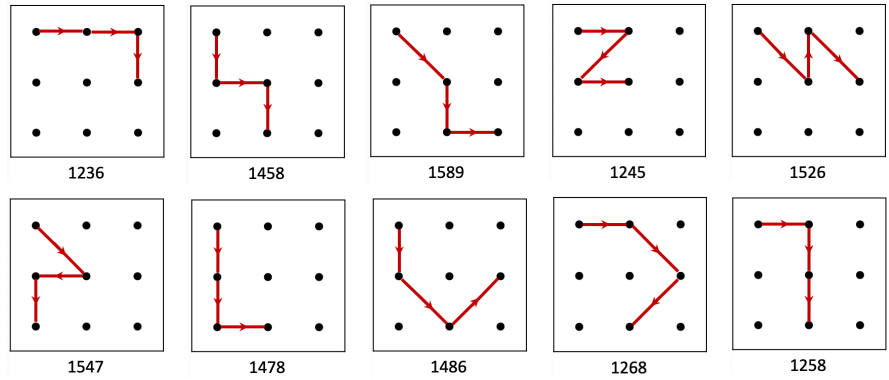}
\vspace{-36mm}
\caption{Fourgram patterns on a 3 $\times$ 3 grid and their numerical representations, used in our implementation.}
\label{fig:patterns}
\end{figure}

\subsubsection*{PT-2D-2FA}  
We assume the user owns a device with a touchscreen or a keypad but not a camera (e.g., a smartwatch). 
For such devices we recommend representing \ids as \textit{patterns}, similar to the one used in Android pattern lock. 
Several studies have shown convenience of this method 
\cite{von2013patterns,schloglhofer2012secure,andriotis2016study,micallef2015aren}. 
Pattern lock screen is a n $\times$ n grid of dots on which the 
user can draw patterns by connecting some number of dots with straight directional lines. 
We can interpret a pattern numerically too
by assigning a number to each dot. Therefore, users can also input \ids using a simple 
keypad, which could be helpful in case their device  only comes a keypad and not a touchscreen.




To reduce pattern complexity, we selected fourgram patterns on a 3x3 grid
with the upper-left corner as the starting point to  
match the user's preferred starting point \cite{tupsamudre2018tinpal,loge2016user}. 
To minimize false accepts and false rejects, patterns in the dictionary were selected such that the distance 
between any two is at least 2, i.e., at least two connecting lines differ between them.
In Appendix \ref{sec:patternchoice}, we provide more details on choosing the right patterns.

Figure \ref{fig:patterns} shows the collection of patterns used in our implementation and evaluation, and their numerical representations 
with upper-left dot labeled as 1 and the lower-right dot labeled as 9.



\subsubsection*{QR-2D-2FA}
In addition to touchscreen or keypad, a device may also have a camera using which one can scan \ids encoded as QR Codes. Such a device can naturally support patterns and digits as well. 
QR Code scanning is a semi-automated input method (user only scans the code) and is therefore efficient and has a low error rate. QR Codes are extensively used in 2FA applications for enrollment and authentication. Many 2FA systems encode 2FA secret in a QR Code and ask users to scan it using their phone during the enrollment \cite{qrqws,qrzoom,qrduo}. Others have used QR Code as part of the authentication process. For example, in \cite{ndss-2014}, a challenge message is encoded as a QR Code to be transferred to the device from the client. The PIN is computed as a function of the challenge and encoded into QR Code to be scanned by the client. \nsha{Didn't get this}  OpTFA \cite{jareckitwo} uses QR Codes to establish an authenticated channel between the device and the client as part of the authentication process.  Phoneauth \cite{czeskis2012strengthening} suggests  QR Codes to transfer session information from the client to the server through the device as part of their recovery process. 



\subsection{Server-side Scripts}

The server runs a web service and keeps a database table (named \texttt{users}) of users' credentials, including username, password, and 
2FA secret key. The database also maintains a table of authentication sessions (named \texttt{active\_sessions}) that records 
the issued $\pt$, timestamp, and authentication status flag for each session.

For each authentication attempt, the web page receives the username and password from the user  and navigates her to a 
second web page displaying one of the ``available identifiers'' in the dictionary\footnote{The secondary authentication factor is displayed regardless of the success or failure of 
password authentication to resist password guessing attacks.}. 
For each user, available identifiers are those identifiers in the dictionary that are not in use by the 
current active sessions 
or had been communicated for previous completed authenticated session, but are not yet expired. 

After communicating $\pt$, the server makes a record entry in \texttt{active\_sessions} table 
flagging  $\pt$ as \texttt{active}. 
The server starts a timer \texttt{session\_timer} for the active session 
and waits to receive a PIN and a $\pt$. 
The session will terminate if  \texttt{session\_timer} expires. 

On receiving the PIN and $\pt$,
the server computes the PIN and compares it with the one received. \shashank{Pattern needs to be used to locate the right session.}
 If the received PIN and password are correct, the server locates the right session corresponding to the received $\pt$ and 
authenticates the user by displaying a success message. 
If either the password or PIN are incorrect or $\pt$ does not match any of the 
identifiers issued, the server would abort the protocol and displays an authentication failed message. 

Once the authentication succeeds, 
fails, or times out, the server flags $\pt$ in the \texttt{active\_sessions} table 
as \texttt{succeeded, failed, or timed-out} and starts a $\pt$ expiration timer \texttt{identifier\_timer}. 
The identifier is removed from the  
\texttt{active\_sessions} when \texttt{identifier \_timer} expires and will be available for reuse. 

\subsubsection*{Timer}
A few notes on the timer:
\begin{itemize}
\item Each PIN is generated for a time-slice of 30 seconds as per RFC6238 suggestion.
\item Each PIN is validated on the server against the current time-slice and four further validations for two time-slice backward and two time-slice forward to allow for clock drifts between the device and the server. 
\item \texttt{session\_timer} starts on the server as soon as an identifier is communicated. On expiration of the timer the authentication fails. This timer limits the time  a user can complete an authentication session. This timer can extend to few minutes for usability purposes. There is no security implication on the length of the timer. This timer is set only to make sure an identifier is not occupied indefinitely. 
\item \texttt{identifier\_timer} starts on the server as soon as an authentication is completed (succeeded, failed, or timed-out). On expiration of the timer the identifier is removed from the \texttt{active\_sessions} table  and is available for reuse. This timer ensures that a identifier is not repeated early enough for an attacker to replay a valid PIN.
\end{itemize}

\hit{
\subsection{2FA Android Application}

We developed an Android application for both PT-2D-2FA and QR-2D-2FA variants.
The registration process
is similar to that of PIN-2FA, in which the user
inputs the 2FA secret associated with an account (username@
server), either manually or by capturing a QR
Code. During authentication, the user selects the account
(if more than one account is registered on the 2FA app),
and then inputs the identifier.
In  PT-2D-2FA,  a 3 $\times$ 3 dot grid is shown on the app where 
the user can draw the pattern by connecting the dots with straight lines. The pattern is received by 
the application and is encoded into digits by assigning a number to each dot. 
In QR-2D-2FA, the app shows a QR Code scanning page. The user can scan the identifier displayed on the webpage as a QR Code. 
The QR Code is decoded and passed to PIN generation function. We used Google Mobile Vision Barcode API to deploy  QR Code scanning functionality.

The PIN generation function implementation has forked a simple implementation of  TOTP in Java  \cite{2facode} 
with several modifications to allow the use of other hash functions and PIN length, and 
to incorporate the pattern in the PIN computation. 
PIN along with the pattern is submitted to the server as an HTTP POST request. 


Our approach assumes that the device can connect to the server (e.g., over the internet).  But in case the device loses connectivity, we can fall back to a \stfa\ like method. After the user inputs the \id on the device, the computed PIN will be displayed to her. She can manually copy the PIN to the client terminal, which would then be transferred to the server. }

\subsection{PIN Generation \& Validation Function}

Our implementation of the PIN generation function is developed on top of TOTP  \cite{totp}
 with several modifications as will be discussed here. 
We compute HMAC-SHA256
of  $\pt$ concatenated with the Unix timestamp (in 30-second time-slice intervals) using a 128-bit 2FA secret key as the HMAC key. 
The HMAC output of size 256bit is considered as the PIN.

To compensate for  possible time drifts between the server and the device, and any 
computation or communication delays, 
the received PIN would be verified by the server if it matches any of the 
 PINs generated on the current time-slice and two time-slices, before and after 
 the current time-slice. 

Some of the design choices in our PIN generation implementation are not in compliance with common industrial deployments 
since their choices  might have been deprecated or other secure 
replacements might be available. 
For example, SHA1 suggested in \cite{totp} had been attacked in theory and practice and is announced deprecated by NIST. 
Similarly, 6-digit PIN was adopted for easier manual copying by the users. Since, \ptfa\ automates the PIN transfer we 
can increase the PIN length. 
For PIN-2FA method evaluated in our usability study  (Section \ref{sec:usability}) we considered the common deployment with SHA1 and 6-digit PIN to match Google Authenticator app. 

\section{Usability Study}
\label{sec:usability}

\subsection{Study Setup}

To evaluate usability of 2D-2FA, we ran a user study with 30 participants.
We studied our PT-2D-2FA and QR-2D-2FA as well the commonly used PIN-2FA for comparison. 
The hypothesis is that 2D-2FA is at least as usable as PIN-2FA while providing higher lever of security and efficiency. 
Although patterns can be read and 
input in their numerical representation, we did not study this due to its similarity to PIN-2FA. 

The study was performed in-person in a controlled environment and the recruitment was through word of mouth. 
Given the COVID-19 pandemic and work-from-home 
situation, we met no more than 2 participants at a time, at their place of choice (examiners' homes, participants' home, or an outdoor area).  
Common safety practices including face masks, social distancing, and sanitization were followed. 
The user study was discussed with our organization's legal, privacy and finance teams to confirm ethical and privacy considerations. 
The study took about 30 minutes to complete for each participant and they were gifted a \$25 amazon card for their time.

A laptop and a phone (same devices mentioned in Section \ref{sec:deploy}) were handed over to the participants. 
The web-service and the database server were running on the laptop. The smartphone had the 2D-2FA app 
and the android version of Google Authenticator app installed. 
Internet connectivity was provided by a mobile hotspot on Verizon network. 
Timing information as well as the 
number of successful and failed attempts were recorded in a database for  performance and error rate  analysis.   
We used Google forms to present the study instructions and questionnaires, and to collect survey responses. All participants were given 
the same information about the study and three 2FA methods. To avoid influencing the participants, we did not disclose which one of the methods were designed by us.

The 2FA enrollment process is the same for all methods. It just involves copying the 2FA secret  from the server to the device (through the client). Therefore, we evaluated the login process only. 
The first page of the study website was a simple username and password webpage. Upon entering username/password, the second step verification webpage was loaded. The second step webpage for \stfa had a text box to enter the PIN. The participants were instructed to enter the PIN as shown on the Google Authenticator app.  
Entering the correct/incorrect PIN navigated them to a success/fail webpage.  
The second step webpage for PT-2D-2FA and QR-2D-2FA displayed a pattern and a QR Code, respectively, and waited for up to 30s
for the PIN and \id to be received from the device before navigating the user to the success or fail webpage. 

Below, we outline the study flow.

\mypara{Consent} First, the participants were informed that we intend to evaluate usability of three 
different 2FA systems. The participants were shown a consent form stating that personally identifiable data will not be 
collected and any recorded data will be kept confidential. Participation in the study was voluntary and the participants 
were given the option to withdraw from the study at any point 
if they did not feel comfortable with it. 

\mypara{Introduction} Participants were shown a short description of the three 2FA methods: PIN-2FA, PT-2D-2FA, 
and QR-2D-2FA, and instructions on how to use them. 

\mypara{Learning} We asked the participants to login to the study website for as many times as needed to get familiar with the three methods.  
To only estimate delay of the 2FA process, the username and password were pre-filled on the website in all the three cases.

\mypara{Main task} Each participant was instructed to complete a login attempt to the study website for 10 times using each of the 
three methods. The ordering of the three methods was randomized to eliminate biases. 

\mypara{Survey Questions} For each of the three methods, we asked a total number of 18 5-point Likert questions. 
Specifically, we presented several statements (both with positive and negative weights) about each method and asked the participants 
to express their agreement/disagreement with the statement.  The list of questions can be found in Appendix \ref{sec:usability-questions}.  
10 out of the 18 statements were the standard System Usability Scale (SUS) \cite{sus} questions 
that measure usability of any given system with respect to ease of use, learnability, and adoptability. The remaining  
8 questions were designed to understand users' perception of efficiency, security, accuracy, adoptability, 
and instructions and feedback.
After these questions, we asked the participants which method they preferred and the reason for their preference. 
At the end of the study, the participants filled out a demographic information questionnaire. 

\subsection{Performance Evaluation}
\label{sec:eval}

\begin{table}
\caption{Execution time of different operations and the overall delay for the studied 2FA methods}
\label{tab:timing}
\centering
\small
\begin{tabular}{|l|l|l|}
\hline
\textbf{Operation}       & \textbf{Party} & \textbf{Avg Time} \\ \hline
\textbf{PIN generation}  & $\Dv$         & 0.03 ms           \\ \hline
\textbf{PIN verification}  & $\Sr$         &  0.00 ms                 \\ \hline 
\hline 
\textbf{Overall PT-2D-2FA}   & $\Dv$, $\Sr$, $\Us$            &   11.87 s                \\ \hline 
\textbf{Overall QR-2D-2FA}   & $\Dv$, $\Sr$, $\Us$            &   5.12 s                \\ \hline 
\textbf{Overall PIN-2D-2FA}   & $\Dv$, $\Sr$, $\Us$            &   4.59 s                \\ \hline 

\end{tabular}
\end{table}

To evaluate the performance of the operations, we executed the PIN generation on the device and the PIN 
verification on the server, and report on the execution time averaged over 100,000 iterations in Table \ref{tab:timing}. 
As shown in the table, the delay incurred by these operations seem to be negligible.

As part of the user study we measured the time it takes to complete the 2FA process, from the 
time the 2FA webpage is loaded to the time the PIN is verified. Therefore, it includes the 
user interaction delay, i.e., copying the PIN on the browser for PIN-2FA, drawing the pattern, or scanning the QR Code, 
as well as any computation and communication delay. The reported results are averaged over 300 attempts for each method. 
The delay for PIN-2FA was on average 11.87s a the standard deviation of 9.99. 
The delay for PT-2D-2FA  was 5.12s with a standard deviation 
of 3.23, and the delay for QR-2D-2FA was slightly less with an average of 4.59s and a standard deviation of 3.10. 
The minimum delay to complete PIN-2FA, PT-2D-2FA, and QR-2D-2FA, was 4s, 2s, and 1s, respectively. 
The maximum delay was 64s for PIN-2FA, and no timeout was considered. The maximum delay was 24s and 
27s for QR-2D-2FA and PT-2D-2FA, respectively, excluding the attempts that exceeded 30s and timed out. \nsha{but that seems partial?}
As expected, PIN-2FA has the highest delay among all for a 6 digit PIN, while both patterns and QR Codes 
are faster even though a longer PIN (256-bit) is transferred to the server.

Friedman statistical test showed that the difference between the delay is significant at $p<0.05$ with a $p$-value $= 3.53 \times 10^{-67}$  and Chi-square $= 306.022$. 
Further, Wilcoxon Signed-Rank test with Bonferroni correction and adjusted significance level of $< 0.05/3=0.017$ showed statistically significantly difference among 
all three pairs, with $p$-value of  $1.66 \times 10^{-58}$ between PIN and patterns, $3.81 \times 10^{-73}$ between PINs and QR Codes, and  $9.18 \times 10^{-7}$  between patterns and QR Codes.

\subsection{Error Rates}
\label{sec:errorrate}

For all the three methods, we recorded  the number of successful and failed attempts. The error rate was 3.4\%, 2.4\%, and 2.2\% for, PIN-2FA, PT-2D-2FA, and QR-2D-2FA, respectively.   PIN-2FA does not consider any timeout and accepts PINs  generated for the current time-slice as well as two time-slices before and after the current one. Therefore, no timeout error was recorded and all errors were related to entering an incorrect PIN. \nsha{again the timeout}  In 2D-2FA methods, for which a 30s time-out was considered, we noted that all errors that occurred in PT-2D-2FA were for those attempts that exceeded 30s. In QR-2D-2FA,  all but one error was related to expiration of the login session, and one error was due to an incorrect decoding of the QR Code (perhaps a failure in the QR Code scanning API). 

Note that the reported errors contribute to False Reject Rate (FRR). 
We define FRR as the rate of errors in entering PIN, drawing a pattern, or scanning a QR Code, 
such that a valid login attempt by the user does not succeed. 
These errors
might affect the usability of the system since the user needs to repeat the protocol. 
An equivalent of this type of error in \ntfa\ is when the user 
taps on the reject button while in fact she is the one who initiated a login.  
It seems that by extending the expiry time of \ids, the error rate could be reduced. 
We may also consider refreshing the \ids and displaying a new \id on the web-page when the session expires (rather than dropping the login attempt completely). 
This would be similar to the PIN being updated on the Google Authenticator app every 30s.

The fact that no error other that time-out occurred in the \ptfa method implies that the False Accept Rate (FAR) of the system is 0\%. FAR is defined as the rate of error in drawing a pattern or scanning the QR Code such that a different but valid \id in the dictionary 
is generated. Since this mistakenly captured \id
might be issued to an attacker's  authentication session, it may lead to acceptance of the attacker's session and impacts the 
practical security of the system. An equivalent of this type of error in \ntfa\ happens when the user 
taps on the accept button when she is not trying to login.

\subsection{System Usability Questions}
\label{sec:sus}

\begin{figure}[t]
\centering
\includegraphics[width=0.85\textwidth]{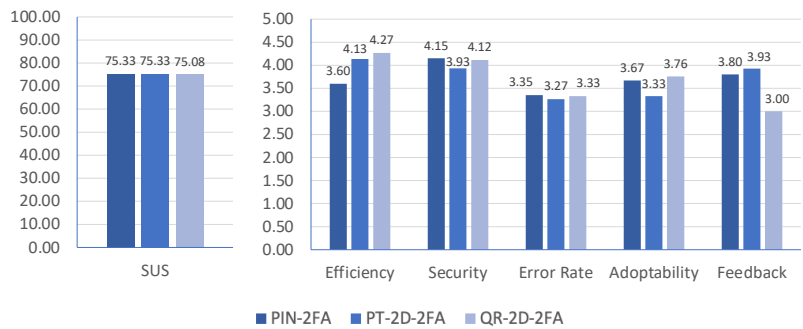}
\vspace{-36mm}
\caption{System Usability Scale and users' perception of Efficiency, Security, Accuracy, Adoptability, and Feedback. \nsha{more space between the figures?}}
\label{fig:usability}
\end{figure}

\mypara{SUS} The average SUS for all the three methods was approximately 75, which translates to a system with "Good" and "Acceptable" usability. 
Precisely the SUS was 75.33 (std. dev. 17.19), 75.33 (std. dev. 13.37), and 75.08 (std. dev. 17.15), for PIN-2FA, PT-2D-2FA, and QR-2D-2FA, respectively. 
Friedman test did not show any statistically significant difference between these methods at $0.05$ significance level, and reported Chi-square $= 4.55$ and $p$-value $= 0.1028$.  The results of this part of the study are summarized in Figure \ref{fig:usability}.

While the SUS was similar among all participants, analyzing individual responses and the open-ended question about the participants' preference showed that 10 participants preferred PIN, 8 participants found patterns to be their method of choice, and 12 participants preferred QR Codes. Importantly, 9 out of 10 participants who preferred PIN mentioned that they have used PIN for several years and are accustomed to it, while patterns and QR Codes are less familiar. Only 1 participant mentioned that she is very comfortable with memorizing numbers, and that is why she prefers this method over any other approach. 

Majority of the participants who liked patterns mentioned that it is easier to remember than a PIN, it is "fun to use", and faster. All the 12 participants who preferred QR Code mentioned the speed and ease of use as the primary factor behind their choice. 
Given that our quantitative analysis shows that pattern and QR Codes have lower error rate and delay compared to PIN, we believe the users would be willing to adopt them if they get familiar with it.  

\mypara{Perception of Efficiency} Answer to 5-point Likert questions  regarding the efficiency of the system (with 5 representing Strongly Agree and 1 representing Strongly Disagree) shows that users were leaning towards "strong agreement" with efficiency of 2D-2FA,  with a score of 4.13 and 4.27 for patterns and QR Codes, respectively. This score was 3.60 for PIN, which shows that users did not feel PIN is as fast as the other two methods. No statistically significant difference was observed among the three methods using Friedman test that resulted in Chi-square $= 5.7167$ and  $p$-value $= 0.05736$.

\mypara{Perception of Security} We asked the participants about their perception of security of the method. The responses show that most participants felt PINs and QR Codes are more secure compared to patterns. The score was 4.15, 3.93, and 4.12, for PIN-2FA, PT-2D-2FA, and QR-2D-2FA, respectively. We informally asked the participants why they would assume that patterns are not as secure as the other two. Some of them felt that patterns are easy for someone else to observe and learn. Note that the participants were not informed about the protocol and the role of patterns. Therefore, they did not know that observing a pattern dose not help an attacker. Also they did not know that the pattern does not need to be complex or long for the system to be more secure. Friedman test did not show any statistically significant difference and resulted in Chi-square $= 2.15$ and $p$-value $= 0.3413$.

\mypara{Perception of Accuracy} In response to questions regarding accuracy of the system and the number of errors, participant were leaning more towards agreement than disagreement that the three methods have low error rate, with a score of 3.35, 3.27, 3.33, for PIN-2FA, PT-2D-2FA, and QR-2D-2FA, respectively. 
Even though the quantitative analysis showed the error rate to be higher for PIN-2FA in practice, the response to this question was more in favor of PIN-2FA. For this score also Friedman test did not show any statistically significant difference with Chi-square $= 0.45$ and $p$-value $= 0.79852$.

\mypara{Adoptability} In response to question about adoptability of the system, we noticed that the participants preferred to use QR Code the most if offered as a login system. The score for this question was 3.67, 3.33, 3.76, for PIN-2FA, PT-2D-2FA, and QR-2D-2FA, respectively. The rating was inline with the answer for the open ended question about preferred method.  The Friedman test showed no statistically significant difference with Chi-square $=  3.65$ and $p$-value $= 0.16122$.

\mypara{Instruction and Feedback} We gave similar instructions about the three methods and the study web-site was the same for them. Our hypothesis was that the answer to the questions regarding the instructions and feedback would be similar for the three methods. The participants were more in agreement than disagreement that the system provided sufficient instructions and feedback, with a score of 3.80, 3.93, and 3.0, for PIN-2FA, PT-2D-2FA, and QR-2D-2FA, respectively. We believe the lower rating of QR Code is perhaps due to the fact that the user interaction was fast and semi-automated (only scanning the QR Code) and therefore the users could not 
notice the feedback on their phone. This is in contrast to PIN and pattern that require more involvement from users, giving them more opportunity to notice the details on the app.  Running the Friedman test showed Chi-square $= 2.7167$ and $p$-value $= 0.25709$, which indicates no statistically significant difference.

\subsection{Demographic Information}
\label{sec:demog}

The demographic information is summarized in Figure \ref{fig:demo}.
10\% of the participants were  55-64 years old, 33\%  were 35-44 years old and 57\% were 25-34 years old. 
53\%  of the 30 participants were male and 47\% were female. 43\% of the participants had a doctorate degree (in various fields, including, medical, art, science, and engineering), 40\% had an MS, MEng, or MA degree, 13\% had a BS or BA degree, and 4\% had a high-school degree. 70\% of the participants had an occupation related to technology,  computer science, or computer engineering, while  30\% were occupied in other domains like art, business, marketing, or healthcare. Majority of the participants declared they have excellent General Computer skills, while 16\%, 10\%, 0\%, and 4\% had good, average, poor, and very poor General Computer skills, respectively. Majority of the participants were familiar with PIN-2FA method, which might explain the SUS Scale of 75.  60\% of the participants were extremely familiar with PIN-2FA, 26\% were familiar with PIN-2FA, and the rest of the participants were equally, somewhat, slightly, or not at all familiar.

\begin{figure*}
\centering
\boxed{
\small
\begin{tabular}{l|l}
Gender                           & \begin{tabular}[c]{@{}l@{}}Female: 47\%, \\ Male: 53\%, \\ Other: 0\%\end{tabular}                                                                                                                                                                                      \\ \hline
Age                              & \begin{tabular}[c]{@{}l@{}}18-24: 0\%, \\ 25-34: 57\%, \\ 35-44: 33\%, \\ 45-54: 0\%, \\ 55-64: 10\%, \\ 65+: 0\%\end{tabular}                                                                                                                                \\ \hline
Education                        & \begin{tabular}[c]{@{}l@{}}Less than high school degree 0\%\\ High school degree or equivalent 4\%\\ Some college but no degree 0\%\\ Associate degree 0\%\\ Bachelor's degree 13\%\\ Master's degree 40\%\\ Doctorate degree 43\%\end{tabular} \\ \hline
Computer Science Professional    & \begin{tabular}[c]{@{}l@{}}Yes 70\%\\ No 30\%\end{tabular}                                                                                                                                                                                                                 \\ \hline
General Computer Skills          & \begin{tabular}[c]{@{}l@{}}Excellent 70\%\\ Very Good 16\%\\ Good 10\%\\ Fair 0\%\\ Poor 4\%\end{tabular}                                                                                                                                                         \\ \hline
Familiarity with 2FA       & \begin{tabular}[c]{@{}l@{}}Extremely Familiar 60\%\\ Very Familiar 26\%\\ Somewhat Familiar 5\%\\ Slightly Familiar 5\%\\ Not at all Familiar 4\%\end{tabular}                                                                                                                                                         
\end{tabular}%
}
\caption{Demographic Information}
\label{fig:demo}
\end{figure*}

\subsection{Study Limitations}
Identifiers could be represented in various
textual and graphical forms. We considered patterns and
QR Codes in the study because they are common and user-friendly. 
One could consider other forms of identifiers too like short strings, etc.
However, we believe that the current study is sufficient to show that \ptfa could offer a good level of usability. Also,
we only considered smartphones as the second factor in the study, since they are more
common compared to other smart devices. One might
consider other form factors depending on the context, like a smartwatch.

\section{Conclusion}
\label{sec:conclude}

We introduced a 2FA method, \ptfa, in which a one-time 
PIN is generated on the secondary factor and is transferred
to the server directly for verification. To associate the PIN received from the device 
to an authentication session, the server sends a unique \id to 
the client, which is displayed to the user on the login terminal and copied by the user to the device. 
The device incorporates the \id in PIN computation and sends it to the server. 
The \id also acts as a user approval of the ongoing authentication attempt. 
On receiving the PIN and the \id, the server can  
authenticate the session for which the \id has been issued by verifying the PIN. 
We define a comprehensive security model and show that \ptfa\ satisfies the security 
goals defined. 

We considered two forms of \ids (patterns and QR Codes), 
developed a proof of concept implementation, and conducted an in-person usability study. 
Our study shows that our system offers a higher accuracy and lower delay compared to 
PIN-2FA method. Our study also indicates that the system usability (SUS = 75) and 
user's perception of efficiency, security, and accuracy is high and comparable to 
the well-known PIN-2FA. Since 75\% of the participants preferred our method over PIN-2FA, 
we believe that not only is our system secure but it would also be adoptable in practice.


\renewcommand{\refname}{\spacedlowsmallcaps{References}} 

\bibliographystyle{unsrt}
\bibliography{all}

\appendix

\section*{Appendix}




\section{Building the Right Pattern Dictionary}
\label{sec:patternchoice}

We should consider several factors in selecting patterns as discussed below.


\mypara{Pattern Length} While the set of patterns could consist of various lengths, we prefer to 
restrict the pattern length to a fixed number to detect drawing mistakes \cite{von2013patterns}, such as  
 addition or deletion of a connecting line. For example, if the pattern length is 4, for a sample pattern ``1234'', 
errors such as ``123'' (aborting) or ``12345'' (addition) would be detected and possibly reversed. 
The length 4, creates 1624 valid patterns \cite{sun2014dissecting}, which is 
a generously large domain size for our application. \shashank{I am curious how this number is calculated. $9 \times 8 \times 7 \times 6 \times 5$ doesn't work.}

\mypara{Pattern Usability Measure} The strength and complexity of a pattern is defined by metrics like pattern length, starting point, direction, symmetry, crosses and knight moves, 
familiar patterns such as alphabet and numbers, etc. 
When patterns are used for authentication, more complex patterns are preferred due to the predictability factor. 
For example, a majority of users select a starting position on the upper row and  a right-to-left/up-to-down 
direction  \cite{loge2016user}. Therefore, these two choices reduces the pattern complexity,
which is not a desirable security characteristic for authentication application. However, in \ptfa, we emphasize  
the usability of the pattern and, therefore, less complex and even predictable patterns are 
preferred. \shashank{as long as it's not easy to make mistakes?}

\mypara{Similarity Metric} While patterns are perceived to be usable 
(please refer to Section \ref{sec:usability}), the error rates are shown to be higher compared to PIN entry 
\cite{von2013patterns,harbach2016anatomy}. Some of these errors reported in prior work are related to 
memorability of patterns (i.e., recalling a previously picked pattern for authentication). Memorability errors are 
not applicable to our application since we present a fresh pattern for each session and ask the 
user to draw the pattern as displayed on the screen. 

However, other type of errors, referred to as slip errors \cite{von2013patterns},  
might not only affect usability of the system by increasing the false rejection, but they may also increase the 
false accepts if 
the user accidentally draws the pattern associated with an attacker's session.
For example, by just one wrong connecting line, pattern ``12365'' would 
be changed to ``12368''. If the two patterns are in the dictionary and one has been issued to the valid user 
and one to the attacker, the user's mistake leads to acceptance of the attacker's session. 

We refer two patterns that can be converted to one another to be ``similar'' and rank the similarity based on the 
number of wrong connecting lines that could convert one to another. 
Therefore, the similarity metric between two patterns is defined as 
the number of connecting lines (insertion, deletion, replacement) needed to map one pattern to another. 
As mentioned earlier some 
mistakes, including deletion or aborting, and insertion can be avoided by fixing the length of the pattern. \shashank{The readers may get a bit confused here. It's not clear how the previous insertion/deletion (in the context of connecting lines) is different from the insertion/deletion here.}
Other errors could be detected if the server avoids concurrently issuing similar patterns (those that may convert to one another 
with one or more errors) by running a similarity check. 
Since we can build the pattern set in advance, inclusion of similar patterns in the pattern dictionary 
can be avoided. 
This would reduce run-time overhead of similarity check on the server.


\shashank{The similarity metric paragraph is quite large (seems to take more than 1/4th of a page). We could consider splitting it up or not making it a bullet point at all. Stepping back a little, may be the four bullet points should not be treated at the same level. We need to talk about each one of them but perhaps the organization could be different.}

\section{Usability Study Questions}
\label{sec:usability-questions}

The response options were Strongly Disagree (1), Disagree (2), Neither agree not disagree (3), Agree (4), and Strongly Agree (5).

\smallskip\noindent \textbf{System Usability Scale.} 
\begin{enumerate}
\item I think that I would like to use this system frequently. 

\item I found the system unnecessarily complex. 

\item I thought the system was easy to use. 

\item I think that I would need the support of a technical person to be able to use this system. 

\item I found the various functions in this system were well integrated. 

\item I thought there was too much inconsistency in this system. 

\item I would imagine that most people would learn to use this system very quickly. 

\item I found the system very cumbersome to use. 

\item I felt very confident using the system. 

\item I needed to learn a lot of things before I could get going with this system.
\end{enumerate}

\medskip\noindent \textbf{Perception of Efficiency.}

I found login to the system to be fast.

\newpage

\medskip\noindent \textbf{Perception of Security.}

I can trust this login method. 

I found the login method
to be secure.

\medskip\noindent \textbf{Perception of Accuracy.}

The number of
unsuccessful attempts
was high.

I found the system to
have a reasonable low
error rate.

\medskip\noindent \textbf{Adoptability.}

If this system is offered
as a login  I would
prefer to use it.

\medskip\noindent \textbf{Instruction and Feedback.}

I found the instructions
to be useful.

The system provided
sufficient feedback when
login attempt was
successful or unsuccessful.

\end{document}